\documentclass[aps,prl,twocolumn,showpacs,superscriptaddress]{revtex4}
\usepackage{graphicx}

\begin{document}

\title{Nature of Correlated Motion of Electrons in the Parent Cobaltate Superconductors}
\author{M.Z. Hasan}\altaffiliation{To whom correspondence should be addressed:
mzhasan@Princeton.edu} \affiliation{Department of Physics, Joseph
Henry Laboratories, Princeton University, Princeton, NJ 08544}
\author{D. Qian}
\author{Y. Li}
\affiliation{Department of Physics, Joseph Henry Laboratories,
Princeton University, Princeton, NJ 08544}
\author{A.V. Fedorov}
\author{Y.-D. Chuang}
\author{A.P. Kuprin}
\affiliation{Advanced Light Source, Lawrence Berkeley National
Laboratory, Berkeley, Ca 94305}
\author{M.L. Foo}
\author{R.J. Cava}
\affiliation{Department of Chemistry, Princeton University,
Princeton, NJ 08544}

\date{\today}

\begin{abstract}
Recently discovered class of cobaltate superconductors
(Na$_{0.3}$CoO$_2$.nH$_{2}$O) is a novel realization of
interacting quantum electron systems in a triangular network with
low-energy degrees of freedom. We employ angle-resolved
photoemission spectroscopy to uncover the nature of microscopic
electron motion in the parent superconductors for the first time.
Results reveal a large hole-like Fermi surface (consistent with
Luttinger theorem) generated by the crossing of super-heavy
quasiparticles. The measured quasiparticle parameters collectively
suggest a two orders of magnitude departure from the conventional
Bardeen-Cooper-Schrieffer electron dynamics paradigm and unveils
cobaltates as a rather hidden class of relatively high temperature
superconductors.

\end{abstract}


\maketitle

Research on strongly correlated electron systems has led to the
discovery of many unconventional states of matter as such realized
in the high temperature superconductors, quantum Hall systems or
in the charge-orbital ordering magnetic compounds. Recently,
attention has focused on layered cobalt oxides
Na$_x$CoO$_2\cdot$nH$_2$O - a new class of doped Mott insulator
which in addition to being strongly correlated are also frustrated
in their magnetic interactions. Depending on the carrier
concentrations, cobaltates exhibit colossal thermoelectric power,
dome-shaped superconductivity and charge-density-wave
phenomena\cite{1}-\cite{5}. They carry certain similarities such
as the two dimensional character of non-Fermi-liquid electron
transport, spin-1/2 magnetism and dome-shaped superconductivity as
observed in the cuprate superconductors\cite{1}-\cite{7}. Unlike
cuprates, cobaltates can be carrier doped over a much wider range
and allow to study emergent behavior in a very heavily doped Mott
insulator. Previous ARPES studies have focused on the high doping
regime (x=70\%, TE-doping) where colossal thermopower is observed
\cite{8}. Transport, magnetic and thermodynamics measurements
suggest that the electron behavior is dramatically different in
low sodium doping (x=30\%, SC-doping) where superconductivity is
observed \cite{5,9}. We have carried out (for the first time) a
detailed microscopic study of the single electron motion in the
parent cobaltate superconductor (low doping regime) class which
reveals the novel many-body state of matter realized in this
system.

Spectroscopic measurements were performed at the Advanced Light
Source. Most of the data were collected with 90 eV or 30 eV
photons with better than 30 or 15 meV energy resolution, and an
angular resolution better than 1\% of the Brillouin zone at ALS
Beamlines 12.0.1 and 7.0.1 using Scienta analyzers with chamber
pressure better than 8$\times$10$^{-11}$ torr. A key aspect of the
success of our spectroscopic studies has been a careful growth and
characterization of high quality single crystals and precise
control of sodium concentration\cite{2, 5}. Cleaving the samples
in situ at 20 K (or 100K) resulted in shiny flat surfaces,
characterized by diffraction methods to be clean and well ordered
with the same symmetry as the bulk. No Ruthanate-like surface
state behavior was observed in high quality samples and all the
data taken at 20K although cleaved at 100K in a few cases for
careful cross-checking.

Fig-1(a-d) shows representative single electron removal spectra as
a function of energy and momentum in Na$_{0.3}$CoO$_2$. A broad
quasiparticle (electron dressed with interactions) feature is seen
to disperse from high binding energies at high momentum values
near the corner (K) or the face (M) of the crystal reciprocal
space (Brillouin zone, BZ) to the Fermi level (zero binding
energy). This electron-feature grows in intensity (yellow emerging
from red) before crossing the Fermi level. This is the
quasiparticle band in this system. It is a hole-like\cite{8,9}
band since the bottom of the band is located near the zone
boundary (K or M) as opposed to the zone center. We trace the
momentum values (k) of the electron (quasiparticle)'s zero-binding
energy crossing point over a large part of the Brillouin zone.
This allows us to generate a momentum space density of states
"n(k)" plot (Fig.2(a)). The inner edge of this density plot is the
Fermi surface. It is a large rounded hole around the zone center.
The Fermi surface of the parent superconductor (x=30\%, SC-doping)
as we report here exhibits more hexagonal character than the
related thermoelectric compound (x=70-75\%, TE-doping)\cite{8}.
This shape is similar to the Local Density Approximation
calculation\cite{10}. However, no satellite pocket is observed.
This could be due to strong correlation effects, which can push
the minority bands away from the Fermi level deep below in binding
energies and wash out their relative intensity. A large Hubbard-U
($\sim$ 4 eV) can be extracted by fitting the valence excitations
with a cluster model \cite{8}. The size of the Fermi surface,
k$_f$ $\sim$ 0.8 \AA$^{-1}$, for the SC-doping studied here is
larger than the size observed in the TE-doping\cite{8} suggesting
a hole doping picture of these compounds (since x=30\% samples
have more holes)(Fig.2(b)). The Fermi surface area suggests that
about 64\% of the Brillouin Zone is electron-occupied. This is
consistent with the conventional electron count in the x=30\%
compound. Hence Luttinger theorem is satisfied in a conventional
sense. For the SC-doping, the carrier sign of thermopower or Hall
transport\cite{4,5} being positive at all temperatures suggests
that it is the holes that carry the current - a fact consistent
with our finding of a hole-like Fermi surface.

\begin{figure}[t]
\center \includegraphics[width=8.6cm]{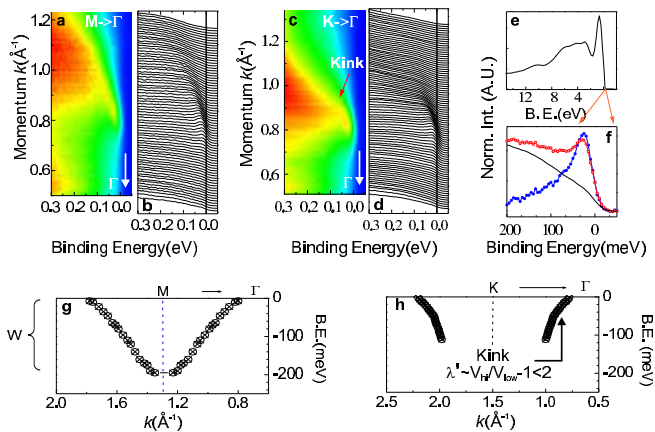}
\caption{\textbf{Single-Particle Dispersion}: Zone face to the
zone center (M$\rightarrow\Gamma$): a, Spectral intensity. b,
Energy dispersion curves; Zone corner to the zone center
(K$\rightarrow\Gamma$): c, Spectral intensity. d, Energy
dispersion curves. Color red/blue corresponds to high/low
intensity. e, Valence band. f, Quasiparticles. g, Dispersion
relation, E vs. k along M$\rightarrow\Gamma$ (extracted after
background correction and symmetrized around the M point). The
bandwidth ("W" $\sim$ 200 meV) is estimated by
tracing/extrapolating the band to the zone boundary shown in (g).
h, The raw dispersion (background not subtracted) along the
K$\rightarrow\Gamma$ cut shows a kink (also seen in the raw data
(c). All the data were taken at 20K.}
\end{figure}

In order to estimate the total bandwidth we need to obtain the
complete dispersion relation (energy vs. momentum) over the full
Brillouin zone. To achieve this we extrapolate the quasiparticle
band to the zone boundary (Fig.1(g)). This gives a value of the
total energy dispersion, or bandwidth, of about 180 meV $\pm$ 20
meV ($\sim$ 0.2 eV). This bandwidth is about a factor of two
larger than the bandwidth observed in the thermoelectric
(TE-doping, x $\sim$ 70\%) cobaltates\cite{8} suggesting a more
delocalization (electron-wavefunctions spread out more) of
carriers in the parent superconductor. Larger bandwidth may be
responsible for weaker thermopower observed near superconducting
doping because within a Fermi liquid picture thermopower is
inversely related to the bandwidth\cite{3,9}. However, the
observed bandwidth is still much smaller than the value calculated
($\sim$1.5 eV) using the Local Density Approximation
methods\cite{10} which does not consider electron correlation
effects. The bandwidth measured here is consistent with that
extracted/estimated from the bulk thermal measurements on the
hydrated superconducting samples\cite{7}. The specific dispersion
behavior (band having maximum at the zone boundary or emanating
from the zone boundary to cross Fermi level in moving towards the
zone center) indicates a negative sign \cite{12}-\cite{14} of
effective single-particle hopping (\textit{t}$<$0). This sign is
crucial in describing the physics in a triangular network. The
possibility of superconductivity and its doping dependence in
triangular many-body models such as Hubbard model or \textit{t-J}
model crucially depend on the sign of \textit{t}
\cite{12}-\cite{15}. Hence, our unambiguous determination of a
negative sign of \textit{t} in the SC-doping compound has
implications for the possibilities of superconducting order
parameters (such as $d_{x^2-y^2}$) it can sustain: $d_{x^2-y^2}$
superconductivity is fragile for a negative \textit{t} model
whereas it is robust for positive \textit{t}. Furthermore, a
negative \textit{t} rules out the possibility of Nagaoka-type
ferromagnetic instability\cite{13} in these systems. However, next
neighbor Coulomb interaction can stabilize superconductivity in
the presence of a negative \textit{t} and restore Resonating
Valence Bond mechanism of superconductivity\cite{15}.

\begin{figure}[t]
\center \includegraphics[width=8cm]{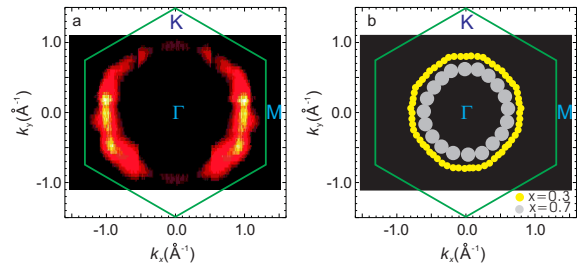}
\caption{\textbf{Fermi Surface of Parent Superconductor}: a,
Momentum-space density of states (n(k)) plot shows a large
hole-pocket centered around the $\Gamma$-point. The inner edge of
the pocket is the Fermi surface which exhibits some hexagonal
anisotropy. b, Comparison of the Fermi surface between parent
superconductor (x=0.3) and the thermoelectric (x=0.7) host
compound.}
\end{figure}

We then estimate an effective Fermi velocity, which is roughly the
slope of the dispersion relation or the derivative of the
band\cite{9} evaluated very close to the Fermi level, by analyzing
the dispersion curves in Fig.1(a-d). The Fermi velocity, averaged
over the entire Fermi surface is found to be about 0.15 $\pm$ 0.05
eV$\cdot$\AA (about 0.2 eV$\cdot$\AA). This is quite small
compared to most known correlated layered oxides. The velocity is
about a factor of 10 smaller than the superconducting
cuprates\cite{16}. Given the size of the Fermi surface we estimate
the carrier mass, m* $\sim$ $\hbar$k$_f$/$|v_f|$ $\sim$ 60 m$_e$.
This is a rather large carrier mass for a high-conductivity
transition metal oxide. Similarly large carrier mass has recently
been reported by muon spin rotation measurements\cite{18}. Such
large carrier masses are typically seen in the heavy-fermion
compounds ($\sim$ 100 m$_e$), where the low-energy physics arises
from two hybridized bands including an atomic-like
\textit{\textbf{f}}-electron band, but are rarely observed in the
transition metal oxides. How cobaltates exhibit such a large
carrier mass where only a single \textit{\textbf{d}}-band is at
play remains to be understood.

We further observe that electron velocity changes beyond 50-70 meV
binding energy range and a kink (Fig.1(c)) is observed in the
dispersion relation along the $\Gamma\rightarrow$K direction in
that energy range. This could be due to coupling to bosonic
collective modes such as phonons or magnetic fluctuations (kink in
Fig.1(c) and 1(g)). The energy scale of the kink - the velocity
cross-over - directly corresponds to the optical phonon
(Raman-active lattice vibrations 55-75 meV) energies in the
cobaltates \cite{19}. Based on the kink parameters (ratio of
high-energy velocity and low-energy velocity) in our data the
effective coupling is on the order of unity which is very similar
to what is observed in the cuprates\cite{16}. A similar kink was
also reported in the high doping regime \cite{8}. Its observation
indicates the existence of strong bosonic interaction or coupling
in the single-particle dynamics of the cobaltates. It would be
interesting to look for such modes using inelastic x-ray \cite{20}
or neutron scattering.

\begin{figure}[t]
\center \includegraphics[width=8.6cm]{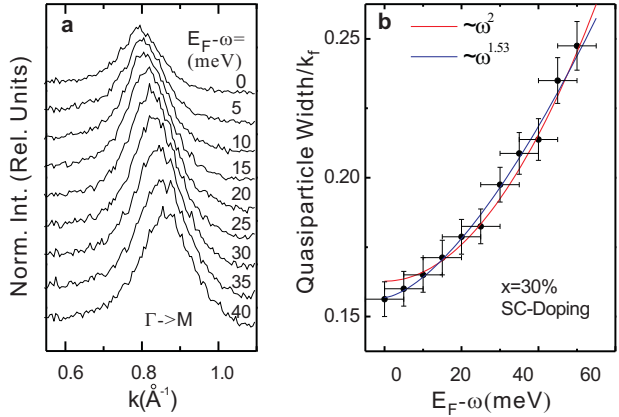} \caption{
\textbf{Self-energy Behavior}: a, Momentum distribution curves
along M$\rightarrow\Gamma$ shows changes of lineshapes away from
the Fermi level (0 meV). b, Quasiparticle (lineshape) width in
x=30\% sample (superconducting doping) plotted as a function of
binding energy, which shows nearly  $\omega^2$ (quadratic)
Fermi-liquid-like behavior.}
\end{figure}

We study the scaling behavior of the electron self-energy based on
the momentum distribution curves shown in Fig.3(a). The
quasiparticle line widths which provide a measure of electron
self-energy or scattering rate, as a function of electron binding
energy ($\omega$) can be fitted with a quadratic-like
($\omega\sim$ 2) form (Fig.3(b)) with a large prefactor than
conventional metals \cite{9,11}. This suggests that the system can
be thought of as a strongly renormalized heavy-Fermi liquid. This
scaling behavior is in significant contrast with our findings in
the thermoelectric cobaltates (x=70\%) \cite{8}.

\begin{table*}
\caption{ARPES Quasiparticle parameters for major classes of
superconductors}
\begin{ruledtabular}
\begin{tabular}{ccccccc}
 Class&T$_c$(K)&Bandwidth&Fermi velocity&Mass
 &Hubbard&Ref.\\
&&W$_{APRES}$(eV)&v$_{f[ARPES]}$(eV$\cdot$\AA)&m/m$_e$&U(eV)&\\
\hline Cobaltates(NaCoO)& \textbf{5} & \textbf{0.18$\pm$0.04} &
\textbf{0.15$\pm$0.10} &
\textbf{60$\pm$20}& {$>$4.0} & \textbf{Present work}\\
\hline
p-Cuprates(LSCO)& 38&$\sim$0.4 &1.8&2 (nodal)&3 - 5&[16] \\
\hline
n-Cuprates(NCCO)&22&$\sim$0.5&2.0&2.4 (nodal)&$\sim$3&[17]\\
\hline Ruthanates(SrRuO) &1&0.5&0.4 (avg)&$\sim$9 (avg)&1&[16]\\
\hline
BCS-type SCs(Pb)&7.2&9.5&12&2& &[21]\\
\end{tabular}
\end{ruledtabular}
\end{table*}

To gain further insights into the nature of electron motion, we
now compare the basic electronic parameters of the cobaltates with
other classes of layered oxide superconductors and conventional
BCS superconductors (Table 1). Based on our finding of electron
behavior, we note that the cobaltates have a large on-site
electron-electron interaction energy (U, measured in a similar
manner described in Ref. [8]) comparable to the superconducting
cuprate family\cite{16} but both the bandwidth and the Fermi
velocity are several fold smaller and the nodal carrier mass is
orders of magnitude larger (Table 1). To compare superconducting
properties, we define a parameter k$_B$T$_c^{max}$/W which
compares two \textit{fundamental} energy scales in an electronic
system. The quantity k$_B$T$_c$ (T$_c$ is the transition
temperature and k$_B$ is the thermal Boltzmann constant) describes
the superconducting condensation temperature/energy scale and W is
the total effective bandwidth that describes the total kinetic
energy available to the electron system. We make a plot, shown in
Fig.4(A), of the quantity, transition temperature as a fraction of
quasiparticle bandwidth, k$_B$T$_c^{max}$/W$_{ARPES}$ (maximum
transition temperature is considered in doped systems) for several
major classes of layered (quasi-2-D materials)
superconductors\cite{16,17}. We then locate the cobaltate class
based on our data. It is evident from Fig.-4(a) that the ratio,
k$_B$T$_c^{max}$/W$_{ARPES}$ ($\sim$k$_B$*5Kelvin/0.2 eV$\cdot$\AA
$\sim$ 0.002) for the cobaltate system is about two orders of
magnitude larger than the strong coupling BCS superconductors such
as tin (Sn) or lead (Pb)\cite{21}. The cobaltate value is rather
close to that observed in the cuprate superconductors or doped
fullerenes\cite{22} where both strong electron-electron and
electron-boson (or phonon) interactions are observed. A
two-orders-of-magnitude departure from the conventional
phonon-only BCS-like value in the cobaltates is a signature for
its electron dynamics to be of unconventional origin in the sense
that phonons alone can not be accounted for superconductivity.
Electron-electron interaction leads to a reduction or
renormalization of W which is a manifestation of the strong
correlation (Mott) effect. This strong electron-electron (Mott
effect) interaction leads to the enhancement of the quantity
T$_c$/W. In order to further test its relevance to
superconductivity in the sense that T$_c$/W reflects a measure of
relative effective interaction strength (from one class to the
other) we plot this quantity against a known independently
measured parameter - the inverse coherence length (1/$\xi$) which
is an inverse measure of the size of the superconducting
electron-pair wavefunction hence a measure of the overall coupling
strength (interactions) from one materials class to the other. In
Fig-4(b) we plot T$_c^{max}$/W$_{ARPES}$ vs. $\xi^{-1}$ and
observe that this is true for most classes of modern
(unconventional) superconductors as well as for the BCS
superconductors suggesting a universal behaviour. Given the slope
of the line ($\sim$ 0.02 nm) and our T$_c$/W ($\sim$ 0.002) data
on cobaltates measured here, we make an empirical estimate of
1/$\xi$ for the cobaltates which is found to be about (0.002/0.02
nm $\sim$ 0.1) inverse nanometer. This agrees remarkably with
recent estimates of coherence length based on magnetic field
measurements within the same order-of-magnitude\cite{6,7}. We
further plot dc conductivity\cite{5,16,22,23} as a function of
quasiparticle bandwidth in conventional and unconventional
superconducting materials, which suggests that cobaltates
(Fig.-4(c)) fall closer to the cuprates or ruthanates but in a
different regime where magnesium diboride belongs to.

The fact that the cobaltate system has a relatively large
transition temperature as a fraction of quasiparticle bandwidth
and small Fermi velocity is rather remarkable. This suggests that
given the total bandwidth or total kinetic energy, the system has
achieved superconductivity at a relatively high temperature
although the actual T$_c$ is rather low in absolute numbers. In
this sense it is a hidden "high temperature superconductor" even
though it may have a very different order parameter symmetry than
the cuprates.

Our results provide the fundamental electron parameters (Fermi
surface topology, sign of hopping, Fermi velocity, quasiparticle
mass, bandwidth etc.) to construct a model hamiltonian for the
cobaltate class. The unusual character of the microscopic electron
dynamics in the cobaltates observed here suggests that any
comprehensive theory of these materials needs to account for the
fact that a relatively large value of transition temperature is
achieved in an extremely narrow band parent material with
unusually heavy and slow moving carriers where a
two-orders-of-magnitude departure from conventional phonon-only
BCS paradigm is observed in the microscopic electron dynamics. The
observed electron dynamics of the cobaltates also point to an
emergent unifying theme in the intrinsic physics of modern
unconventional superconducting materials.

\begin{figure}[h]
\center \includegraphics[width=8.6cm]{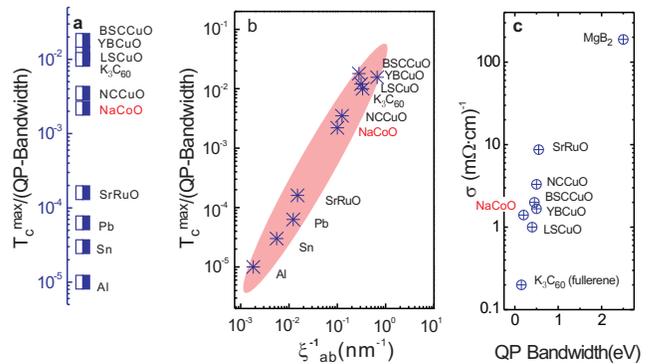}
\caption{\textbf{Quasiparticle-Bandwidth and Superconductivity}:
a, Transition temperature (T$_c$) as a fraction of bandwidth
(T$_c^{max}$/bandwidth) for several classes of superconductors. b,
T$_c$/(ARPES-bandwidth) is found to scale with the inverse
\textit{ab}-plane coherence length. Coherence length data is taken
from Ref. [24]. c. Electrical (dc) conductivity is plotted as a
function of (ARPES) bandwidth for conventional and unconventional
superconductor classes. In general, cobaltate class is found to
cluster with the unconventional superconductors than the BCS type
conventional superconductors. A characteristic
two-orders-of-magnitude departure from conventional BCS-like
behavior is evident in (a), (b) and (c).}
\end{figure}

We greatfully acknowledge N.P. Ong, P.W. Anderson,  G. Baskaran,
P.A. Lee, S. Shastry, S. Sondhi and Z. Hussain for valuable
discussion. Experimental data were recorded at ALS which is
operated by the DOE Office of Basic Energy Science. MZH
acknowledges partial support through NSF-MRSEC (DMR-0213706)
grant. Materials synthesis and ARPES characterization was
supported by DMR-0213706 and the DOE, grant DE-FG02-98-ER45706.

\end{document}